# LASSO extension: using the number of non-zero coefficients to test the global model hypothesis

Authors: Carsten Uhlig (1), Steffen Uhlig (2)
   (1) Munich Biomarker Research Center, Institute of Laboratory Medicine, German Heart Center, Technical University Munich, 80636 Munich, Germany
   (2) QuoData GmbH, Fabeckstr. 43, 14195 Berlin, Germany

## Abstract

In this paper, we propose a test procedure based on the LASSO methodology to test the global null hypothesis of no dependence between a response variable and p predictors, where n observations with $n < p$ are available. The proposed procedure is similar to the F-test for a linear model, which evaluates significance based on the ratio of explained to unexplained variance. However, the F-test is not suitable for models where $p \geq n$. This limitation is due to the fact that when $p \geq n$, the unexplained variance is zero and thus the F-statistic can no longer be calculated. In contrast, the proposed extension of the LASSO methodology overcomes this limitation by using the number of non-zero coefficients in the LASSO model as a test statistic after suitably specifying the regularization parameter. The method allows reliable analysis of high-dimensional datasets with as few as $n = 40$ observations. The performance of the method is tested by means of a power study.

## Introduction

High-dimensional statistical analysis has become a cornerstone of modern scientific research, thanks to the exponential growth of available data and increased computing power. However, as datasets become more complex, new challenges arise. One such hurdle arises when the number of features in a dataset exceeds the number of samples, a situation often referred to as 'p > n'. This challenge, common in emerging fields such as biomarker discovery, can limit the effectiveness of statistical methods and complicate the interpretation of results.

Performing statistical analysis on such high-dimensional datasets can be a significant challenge. Achieving large sample sizes is often limited by high costs and the need to prevent bias introduced by different methods, operators or laboratories (Fan & Lv, 2010; Candès et al., 2018). This complexity is exacerbated by multiple testing, which inflates critical values and risks missing important features (Benjamini & Hochberg, 1995; Dudoit et al., 2003; Zou & Hastie, 2005).

The Least Absolute Shrinkage and Selection Operator (LASSO) technique is known to address some of these challenges by setting irrelevant factors to zero and estimating the relevant ones (Tibshirani, 1996). In this paper, we propose a holistic extension of LASSO. Instead of isolating specific markers, we focus on the total number of features identified by LASSO with non-zero coefficients and aim to evaluate the effectiveness of this method in assessing the relationship between the response and the predictors compared to traditional methods such as Student's t-test.

Our methodology is based on a special technique for determining the regularization parameter, and incorporating Monte Carlo simulation for null hypothesis testing to identify critical values. The paper begins with an explanation of the LASSO based test procedure, followed by a presentation of two power studies: one with a single response-related predictor and another with several relevant features. The paper concludes with a summary of our findings and a discussion, providing insights into assessing the relationship between response and predictors.

# Proposed extension of the LASSO method: testing the global model hypothesis

## Notation

We consider a linear model

$$Y_i = \sum_{j=1}^{p} \beta_j X_{ij} + \epsilon_i$$

for sample (observation) $i = 1, ..., n$, with coefficients $\beta_j$ for the marker level $X_{ij}$ of marker $j = 1, ..., p$. $p$ denotes the number of markers, and $n$ the number of samples, where $p > n$. The error terms $\epsilon_1, ..., \epsilon_n$ are independent random variables with mean 0 and variance $\sigma^2$. This model is the basis of the LASSO method.

If we set

$$Y = (Y_1, ..., Y_n)^T, \beta = (\beta_1, ..., \beta_p)^T$$

and

$$X = (X_{ij})_{i=1,...,n;\ j=1,...,p}$$

the LASSO method can be explained as solving the LASSO functional

$$L = ||Y - X\beta||_2^2 + \lambda ||\beta||_1,$$

where $|| \cdot ||_1$ describes the L1 norm and $|| \cdot ||_2$ the L2 norm. $\lambda$ denotes the regularisation parameter.

If $\lambda = 0$, the LASSO functional simplifies to $L = ||Y - X\beta||_2^2$, and then the solution is equal to the classical multiple linear model. However, a solution is available for $\lambda = 0$ only if $p < n$.

## The global test hypothesis

One of the notable features of the LASSO method is that many of the estimated coefficients $\beta_j$ for the potential explanatory variables $X_{ij}$ in the LASSO model may become equal to 0, and the larger $\lambda$, the smaller the number of coefficients $\beta_j \neq 0$.

A coefficient $\beta_j \neq 0$ does not automatically imply that the coefficient is significantly different from 0. This is due to the fact that the mathematical logic of the LASSO model is based on the minimization of the above functional, but not on a statistical significance test. However, the question whether the null hypothesis $\beta_j = 0$ can be rejected, i.e. whether $\beta_j$ is significantly different from 0, is not the subject of this paper. Instead, we focus on the null hypothesis refered to whether the condition $H_0: \beta_1 = \beta_2 = ... = \beta_p = 0$ holds simultaneously for all coefficients. It is a global null hypothesis, because not a single marker variable is tested, but to check whether there is any (linear) relationship between the influence variables and the response variable.

## The test statistic

This paper proposes to use the number $U = |\{j = 1, ..., p; \beta_j \neq 0\}|$ as the basis of a statistical test for the null hypothesis $H_0: \beta_1 = \beta_2 = ... = \beta_p = 0$. It's important to remember that the result of the LASSO model, and thus $U$, depends on the specific underlying regularisation parameter $\lambda$.

The coefficients of the LASSO model are also affected by the scaling of the marker variables $X_{ij}$. For the purposes of this discussion, we assume that all the marker variables have been linearly standardised, i.e. the mean has been subtracted and the result divided by the standard deviation. This standardisation ensures that the expected value of $X_{ij}$ for fixed j is 0 and the variance is 1. We further assume that the $X_{ij}$ for each fixed j=1,...,p can be considered as realisations of a normally distributed random variable.

In practice, different markers $X_{ij_1}$ and $X_{ij_2}$ for $j_1 \neq j_2$ often have some degree of correlation, which could significantly affect the distribution of $U$ and, consequently, the test result. The question therefore arises how these correlations can be included in the test procedure.

A possible approach to eliminate the correlation between different marker variables is to linearly transform the marker values $X_{ij}$ using the spectral composition of the underlying covariance matrix. The LASSO procedure could then be applied to the resulting random vectors. However, then the resulting LASSO model would presumably include (almost) all marker variables, which is certainly not reasonable. For this reason, this approach is not pursued further in this work.

Computing the LASSO model using the original data rather than transformed uncorrelated data seems preferable. In order to check whether the determined LASSO model with a more or less large number of non-zero coefficients can actually be considered "statistically significant", the observed correlation of the influencing variables must be taken into account when determining the critical values for this test statistic $U = U_\lambda$.

This test variable $U_\lambda$ is intended to test the null hypothesis $H_0: \beta_1 = \beta_2 =...= \beta_p = 0$. This null hypothesis is also tested by the classical F-test for linear models. In contrast to the latter, however, the test proposed here also allows for application in the case that the number of influencing variables is greater than the number of observations. The associated critical value $u_{\lambda,\alpha}$ for $U_\lambda$, above which the null hypothesis must be rejected at a given significance level, can only be determined by means of simulation: Using simulated standard normally distributed dependent variables $Y_i$ and simulated standard normally distributed influence variables $X_{ij}$, the LASSO model is calculated at least 10000 times in order to determine the $(1-\alpha)$ quantile $u_{\lambda,\alpha}$ for $U_\lambda$. In this simulation it is assumed that there is no linear relationship between $Y_i$ and the $X_{ij}$, i.e. $\beta_1 =...= \beta_p = 0$, whereby it is assumed without restriction of generality that the mean value of the $Y_i$ is 0, this in contrast to the above-mentioned F-test, where all parameters are equal to 0 except for one constant. The significance level $\alpha$ is assumed in this paper to be $\alpha = 0.05$.

In the simulation also the correlation between the marker variables has to be taken into account. If $\Sigma$ denotes the covariance matrix of the marker variables, there is an orthogonal matrix $O$ and a diagonal matrix $D$ with non-negative entries so that $\Sigma = OD^2O^T$. These matrices can be used to transform the iid standard normally distributed vectors of marker variables $Z_i = (Z_{i1},...,Z_{ip})^T$, $i = 1,...,n$, to obtain $n$ vectors of correlated marker variables $X_i = ODZ_i$, $i = 1,...,n$. The covariance matrix of these simulated vectors is again equal to $\Sigma$.

## Determination of the regularisation parameter λ

The test procedure proposed here is not based on a continuously distributed test variable, but on a test variable with a discrete probability distribution, which is, however, dependent on the respectively selected regularisation parameter $\lambda$. For many $\lambda$'s there is no exact $(1-\alpha)$ quantile for $U_\lambda$ under the null hypothesis. Therefore $\lambda$ has to be selected by means of simulation calculations so that the condition for a significance test at significance level $\alpha$ is actually fulfilled: $P(U_\lambda > u_{\lambda,\alpha}) = \alpha$ under the null hypothesis. If $P(U_\lambda > r) = \alpha$, the respective regularisation parameter is denoted $\lambda=\lambda_r$, and the corresponding test statistic is denoted $U_\lambda = U_{\lambda_r} = U(r)$, where $r=0,1,2,...$ . $U(0)$ denotes the number of non-zero coefficients in the LASSO model for $\lambda=\lambda_0$, i.e. $\lambda_0$ has to be determined by means of simulation so that $P(U_{\lambda_0} > 0) = \alpha$. Similarly, $U(1)$ denotes the number of non-zero coefficients in the LASSO model for $\lambda=\lambda_1$, i.e. $\lambda_1$ has to be determined by means of simulation so that $P(U_{\lambda_1} > 1) = \alpha$.

$U(0)$ has a very simple interpretation: As long as none of the coefficients of the LASSO-Modell becomes non-negative, the null hypothesis cannot be rejected. However, as soon as one of these coefficients is no longer 0, the null hypothesis H0 must be rejected, i.e. one can directly see from the result of the LASSO model - for the corresponding regularization parameter - whether the model is "statistically significant".

## Overview of the test procedure

The test procedure described here is computationally very complex and is summarized here once again:
1. First, the marker variables are put into standardized form, based on mean and the standard deviation per marker.
2. Then the $p \times p$ covariance matrix of the marker variables is determined and the spectral decomposition of this matrix is performed.
3. The generation of the simulated marker values is based on the spectral decomposition, where the goal is to determine the desired regularization parameter(s) $\lambda_r$ for appropriate $r$.
4. Finally, the calculation of the LASSO model for the actual data is performed using the regularization parameter $\lambda_r$. $U(r)$ represents the number of non-negative coefficients in the respective LASSO model.
5. The null hypothesis is rejected if $U(r)$ is greater than $r$.

## Power study

Whether a proposed statistical test procedure is useful or not depends on its power, i.e., on the probability that the test procedure rejects the null hypothesis under certain assumptions about the relationship between response variable and markers. In the simplest case, there is a linear relationship with exactly one of the $p$ markers, while all other markers are unrelated to the response variable. This is the first of the two scenarios examined below. In the second scenario, it is assumed that not one marker but several markers have a linear and more or less pronounced relationship with the response variable. For the remaining markers, it is again assumed that there is no relationship at all with the response variable. In the power study, we assume $n = 40$ samples and $p = 200$ markers in both scenarios. For simplicity, we assume that the marker variables and the error variable are based on standard normally distributed random variables, i.e. $\sigma = 1$. Furthermore, we assume that there is no correlation between the markers.

For comparison, we also calculate the power of the t test in the simple linear model for these scenarios, which is applied in parallel to all 200 markers. The F-test for the multiple linear model, which would be actually more suitable, cannot be used because of $p > n$. In order to take the influence of the multiple testing into account, we use the Bonferroni correction and, alternatively, the Benjamini-Hochberg method.

10000 simulation runs were used in each setting.

For scenario 1 the following settings were considered: $\beta_1 = 0.1, 0.2, ..., 1$ and $\beta_2 = \beta_3 = ... = \beta_p = 0$.

For scenario 2 the following settings were considered: For $k = 1, 2, 5, 10, 20, 50$ of the $p = 200$ markers, it was assumed that there is indeed a relationship to the response variable. The first $k$ elements of β are drawn from a Gaussian distribution with mean µ and standard deviation $0.5\mu$. The remaining $p - k$ elements are zeros. This implies that

$\beta_1 \sim N(\mu, 0.5\mu^2)$, $\beta_2 \sim N(\mu, 0.5\mu^2)$, ..., $\beta_k \sim N(\mu, 0.5\mu^2)$ and $\beta_j = 0$ for $j = k + 1, ..., p$.

This approach was chosen in order to obtain conditions as close to reality as possible.

Two different values were set for the simulation parameter µ, $\mu = 0.2$ and $\mu = 0.4$.

It should be noted that Scenario 2 in any case involves a mixture of different combinations of betas, i.e., strictly speaking, the power given is not the power for a particular scenario, but an average value across conditions in which different markers have varying degrees of relationship to response.

# Results

## Single Relevant Predictor (k=1)

In the $k = 1$ scenario, the performance of our proposed LASSO extension (U0) was compared to that of the adjusted t-tests and is displayed in Fig. 2. As the intensity of the simulated β (µ) increased, we observed that the power of $U(0)$ consistently improved. This was demonstrated as the power of $U(0)$ increased steadily as mue progressed from 0.3 to 0.4, reaching a peak of 0.9 when mue was equal to 1.

Interestingly, although the higher $U$ values ($U(1)$ to $U(20)$) showed a similar upward trend with increasing mue, their overall power was lower than that of $U(0)$, indicating that $U(0)$ generally outperformed the higher $U$ values as the intensity of the simulated beta increased.

There is a simple explanation for this behavior: Since only one marker is relevant in the simulation model, the LASSO model will usually also identify only this one marker. It will therefore not be very likely that other non-zero coefficients will be identified in addition to $β_1$, i.e. exceeding the value 1 is not very likely for all ($U(0)$ to $U(20)$), and therefore the power for $U(1)$ to $U(20)$ in this case is lower than for $U(0)$.

On the other hand, it should be noted that in this particular case the classical t test for the simple linear model also performs very well. In comparison, the proposed extension of the LASSO procedure is no better. Thus, if one already knows that there is only one relevant marker, it is not worth to replace the traditional t test by our proposed LASSO extension. Nevertheless, the choice between our proposed LASSO extension and traditional t-tests should be based on the specific parameters and requirements.

Also noteworthy is the fact that the t test produces very similar results using both the classical Bonferroni correction and the Benjamini-Hochberg approach. The Benjamini-Hochberg approach was developed for testing single parameters and obviously cannot show its superiority over the Bonferroni approach when testing a global model hypothesis.

## Multiple Relevant Predictors (k >= 1)

In this scenario, we examined situations where there were several relevant predictors. The intensity of the simulated β coefficient (µ) and the number of relevant predictors ($k$) were varied to determine their influence on performance.

Our simulations highlighted several key observations. For a lower µ value of 0.2 (Fig. 1), $U(0)$ consistently delivered the best performance over a range of $k$ up to $k = 20$. However, as the intensity of the simulated β (µ) increased to 0.4 (Fig. 1-C), the relative performance of the different $U$ values varied. $U(0)$ continued to outperform higher $U$ values up to $k = 10$. However, for larger numbers of relevant predictors ($k > 10$), $U(1)$, $U(2)$ and $U(5)$ started to show superior performance.

As the number of relevant predictors increased, the performance of the t-test adjusted for multiple comparisons plateaued and even declined. This is probably due to the limitations of univariate testing methods in dealing with multiple relevant predictors, which increase the unexplained variance. In contrast, the power of our proposed LASSO extension showed an increasing trend with the increasing number of relevant predictors, especially when these predictors had a significant impact on the response variable.

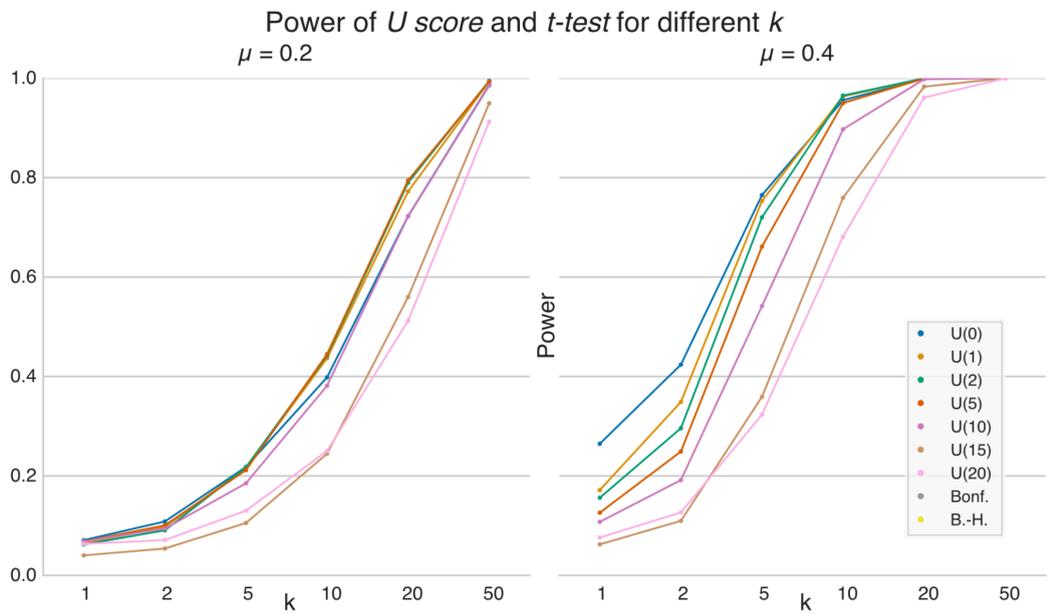

Figure 1: Two power graphs are displayed rendering the behaviour of changing parameter $k$ ($k \geq 1$) with $\mu = 0.2$ and $\mu = 0.4$ towards the power displayed on the y-axis. On the x-axis the number of relationships is evaluated for. U(number) represents the critical threshold at $u_{\lambda,\alpha}$ which means that by testing for $U(0)$ we simply need to check if the chosen model exhibits more non-zero coefficients than none. Bonf. and B.-H. stand for *Bonferroni* correction and *Benjamini-Hochberg* correction.

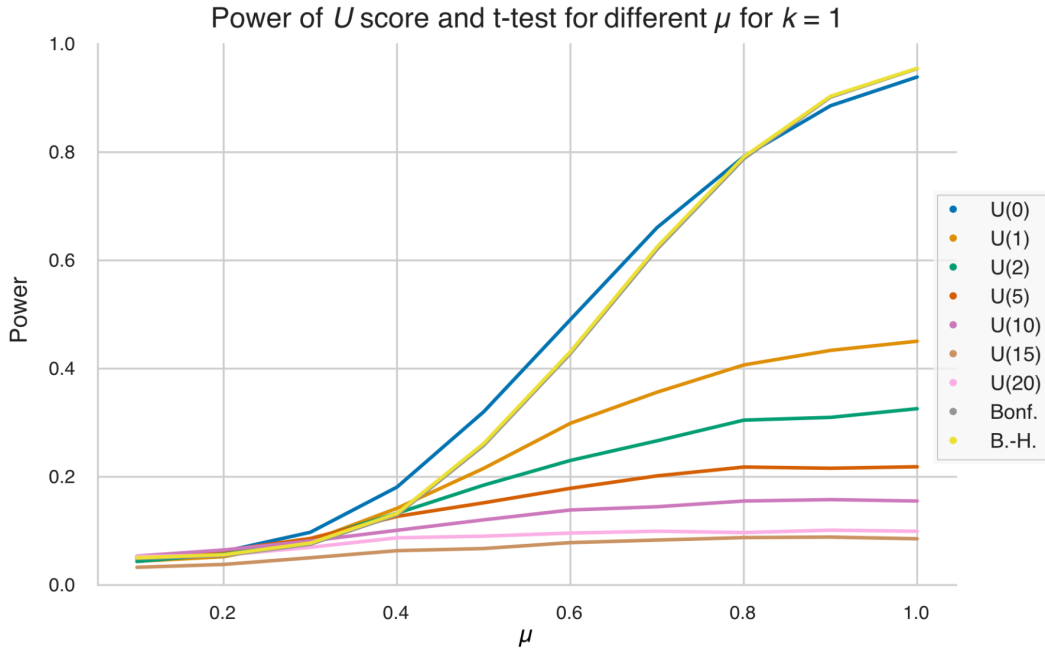

Figure 2: The line diagram presents the scenario for $k = 1$ meaning only one variable is relevant. The intensity for this one variable is evaluated in comparison with the power and then compared between different $u_{\lambda,\alpha}$ and <t-tests. U(number) represents the critical threshold at $u_{\lambda,\alpha}$ which means that by testing for $U(0)$ we simply need to check if the chosen model exhibits more non-zero coefficients than none. Bonf. and B.-H. stand for *Bonferroni* correction and *Benjamini-Hochberg* correction.

# Discussion

To conclude, our proposed extension offers an innovative solution for analyzing high-dimensional datasets with limited samples, especially in scenarios where $p > n$ and traditional test statistics such as the F-test are no longer applicable, and where $n$ is too small for other more sophisticated approaches. Although our extension is global in its approach, it can still provide insightful indicators regarding the number of significant features present in the dataset. When compared to traditional t-tests, our method demonstrates comparable performance for datasets with a single relevant feature. Its true strengths, however, are revealed in multivariate analyses where datasets contain multiple relevant features, showcasing the method's adaptability and efficacy in complex data scenarios.